\documentclass[prb,twocolumn,amssymb,showpacs,superscriptaddress]{revtex4}
\usepackage{graphicx,amsmath,epstopdf,color,ulem,gensymb}

\begin{document}
\title{Multiorbital kinetic effects on charge ordering of frustrated electrons on the triangular lattice}
\author{C. F\'evrier, S. Fratini, A. Ralko}
\affiliation{CNRS, Inst NEEL, F-38042 Grenoble, France}
\affiliation{Univ. Grenoble Alpes, Inst NEEL, F-38042 Grenoble, France}
\date{\today}
\begin{abstract}
The role of the multiorbital effects on the emergence of frustrated electronic
orders on the triangular lattice at half filling is investigated through an
extended spinless fermion Hubbard model. By using two complementary approaches,
unrestricted Hartree-Fock and exact diagonalizations, we unravel a very rich
phase diagram controlled by  the strength of both local and off-site Coulomb
interactions and by the interorbital hopping anisotropy ratio $t'/t$.  Three
robust unconventional electronic phases, a pinball liquid, an inverse pinball
liquid, and a large-unit-cell $\sqrt{12} \times \sqrt{12}$ droplet phase, are
found to be generic in the triangular geometry, being controlled by the band
structure parameters.  The latter are also stabilized in the isotropic limit of
our microscopic model, which recovers the standard SU(2) spinful extended
single-band Hubbard model.
\end{abstract}


\maketitle

\section{Introduction}

In analogy with frustrated spin systems,  frustration of the charge
interactions by the triangular-lattice geometry constitutes a favorable
playground for the emergence of novel phases.  In this context, the pinball
liquid (PL) was proposed a decade ago \cite{Hotta06} as an original
Coulomb-induced charge-ordered metallic phase in the framework of the extended
Hubbard model. This phase  has no classical equivalent and takes advantage of
quantum fluctuations in order to lift the massive degeneracy of interacting
electrons on quarter-filled triangular lattices in the classical limit.  While
the existence of the PL is now a consolidated fact, as it has been demonstrated
by several complementary techniques and approximations
\cite{Kaneko06,Miyazaki09,Cano10,Cano11,Merino13,Tocchio14}, such an interesting
electronic phase has not been observed experimentally in the
$\theta$-(BEDT-TTF)$_2X$ materials for which it was originally predicted.  This
can be attributed to the presence of other competing effects not considered in
the idealized theoretical descriptions, most notably deviations from a
perfectly isotropic triangular lattice, the interaction with the lattice
degrees of freedom \cite{Udagawa07} and the presence of long-range tails of the
Coulomb repulsion beyond nearest neighbors  \cite{Kuroki09,Vlad15}, which
all favor insulating stripe-ordered states.  

There exist other classes of materials with layered triangular lattices and
sizable electronic interactions which do present interesting  charge-ordered
phases whose origin is not fully understood. These include transition-metal
oxides such as the layered cobaltates Na$_x$CoO$_2$, which exhibit complex
electronic patterns which can be tuned by electron doping
\cite{Takada,CavaAlta,Julien2008,Alloul2012}, and the triangular nickelates
AgNiO$_2$ \cite{Coldea07,Coldea11} and Ag$_2$NiO$_2$ \cite{Yoshida2006},
which show a threefold  ordered metallic phase with anomalous metallic
properties.  Another interesting class is that of transition-metal
dichalcogenides.  In $1T$-TaS$_2$, for example, the ordered state displays a
marked Mott character induced by charge modulations with a large periodicity of
$\sqrt{13}\times\sqrt{13}$ \cite{Perfetti06,Sipos}, and various other
periodicities are found in other compounds.  What all these materials have in
common is that electrons live in bands constructed from $d$ atomic orbitals.
Bridging the ideas of frustrated charge order from their initial domain of
application (single-band, layered organic conductors) to such $d$-electron
compounds requires us to account for the presence of multiple bands and to move
to electron densities not restricted to one-quarter filling.

Multiband effects come in two different kinds.  The first  is related to
interactions that are present within the $d$-electron manifold already at the
atomic level, most notably the on-site Hund interactions acting on the magnetic
degrees of freedom \cite{Kanamori}. These are known to favor the emergence of
high-spin states and  have been shown  to strongly enhance the effects of
electronic correlations \cite{deMediciPRL11,deMediciPRB11,Georgesrev}.  Their
ability to stabilize a PL phase with unconventional metallic properties at a
filling of one electron per site has been explored very recently
\cite{Ralko15}.  The second type of multiband effect, which can also lead to
novel and original properties, is of kinetic origin and has to do with the
microscopic form and symmetry of the interatomic electron transfers.  One
remarkable example is the hidden kagome symmetry and flat bands which have been
pointed out in the layered cobaltates \cite{Maekawa}, and which could be
related with the experimentally observed Kagome order in these compounds
\cite{Alloul09,Alloul2012}.  

The purpose of this paper is to  study  how such multiorbital kinetic effects
influence the frustrated charge order on the triangular lattice.  To this aim
we employ an atomistic tight-binding description \cite{SlaterKoster} and set up
a two-orbital extended Hubbard model where the band structure can be tuned
systematically as a function of the microscopic transfer parameters.  Since our
main focus is the exploration of novel charge ordered phases, it is a good starting point 
to resort to a spinless electron description. This approach 
has been shown to capture the correct
ordering patterns realized in 
single-band models \cite{Hotta06,Cano11} in the limit of strong local Coulomb repulsion, 
where the magnetic energy scales are typically 
much smaller than the ones controlling charge ordering.
It has also been successfully used
to study charge ordering in the context of multiband models for correlated
oxides \cite{Jackeli2008} in the ferromagnetic state.

Our results, obtained here at half filling via unrestricted Hartree-Fock (UHF)
mean-field theory and  exact diagonalization (ED) on small clusters, show the
emergence of a rich panel of charge and orbitally ordered phases. Most notably,
we find three robust unconventional charge-ordered phases, whose occurrence can
be tuned  by varying the multiorbital band structure parameters.
	Two of these phases have peculiar properties since a fraction of the electrons forms a charge order with a threefold symmetry breaking, while the other fraction is free to move on the remaining sites of the lattice, forming a honeycomb structure. These phases are called the pinball 
liquid (originally found in quarter-filled lattices and obtained here at
half filling)
and
 the inverse pinball liquid (which can be viewed as the dual to the
PL). The third unexpected phase found in this work is the large-unit-cell $\sqrt{12} \times \sqrt{12}$ droplet phase, also found in the isotropic limit where our model
reduces to the spinful extended Hubbard model on the triangular lattice, where
it was overlooked in previous studies.
  Such phases could be of  relevance to a
variety of triangular $d$-band electron systems such as the cobaltates,
nickelates and dichalcogenides.

This paper is organized as follows. The microscopic model and the two different
methods of solution are described in Sec. II along with their respective
advantages.  The phase diagrams obtained from both methods upon varying the
interaction parameters and the multiorbital band structure parameters are
presented in Sec. III, together with a detailed description and
characterization of the different  ordered phases. Our main results are
summarized in Sec. IV. 

\section{Model and methods}
\subsection{Spinless two-orbital extended Hubbard model}
\begin{figure}[h]
   \centering 
\includegraphics[width=8cm]{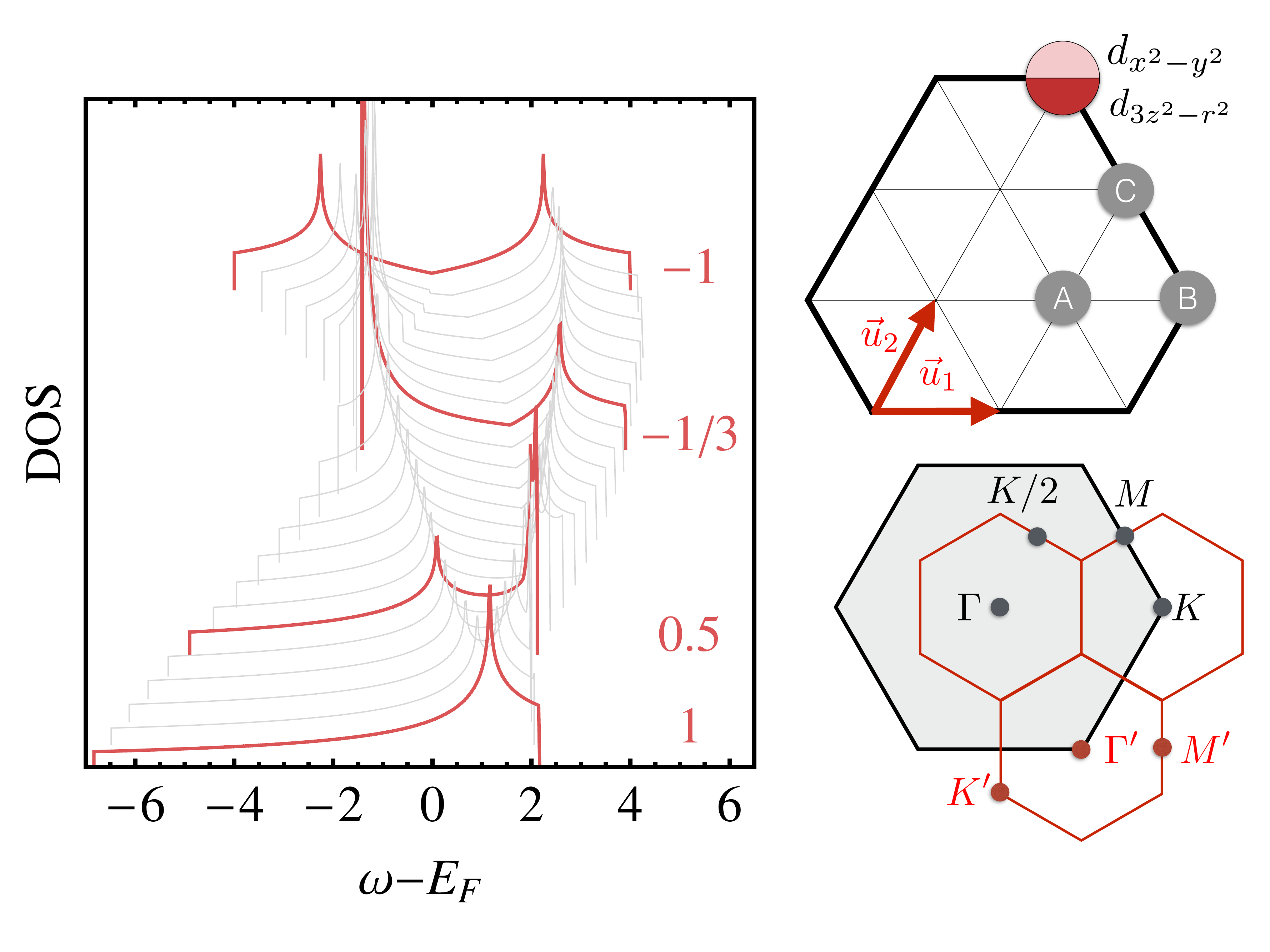}
   \caption{
Evolution of the density of states of the noninteracting system with respect
to the $t'/t$ ratio.  The sketches illustrate (top) the triangular lattice
with two orbitals per site and sublattices A, B and C,  corresponding to
threefold charge ordering, and (bottom) the main symmetry points in the original
and reduced Brillouin zone for three-sublattice ordering.}
 \label{fig:dos}
\end{figure}

\label{sec:model}
To explore the influence of multiorbital effects on Coulomb-driven charge
ordering on the triangular lattice we write the following spinless two-orbital
extended Hubbard model: 
\begin{eqnarray}
H &=& -\sum_{\langle ij\rangle, \tau \tau'}t_{ij}^{\tau \tau'} d_{i \tau }^{\dagger} d_{j \tau'}+\text{H.c.} +  \tilde{U} \sum_{i} n_{i \uparrow} n_{i \downarrow} \nonumber \\
&+& V \sum_{\langle ij\rangle} 
\left(  n_{i \uparrow} + n_{i \downarrow} \right) \left(  n_{j \uparrow} + n_{j \downarrow} \right), \label{eq:H}
\end{eqnarray}
which we study at a density of one electron per site.  The first term describes
$d$-electrons (creation and annihilation operators $d_{i \tau }^{\dagger}$ and $d_{i \tau }$) moving with transfer
integrals  $t^{\tau\tau'}_{ij}$ which depend on both the orbital type and on
the orientation of the bond $(i,j)$ on the triangular lattice. The second term
is an effective on-site Hubbard repulsion between electrons on different
orbitals, with $n_{i,\tau}=d^{\dagger}_{i\tau}d_{i\tau}$ being the local electron density on
orbital $\tau$.  The third term is the Coulomb repulsion between electrons on
neighboring sites, which constitutes the driving force for charge ordering.
Note also that in the electrostatic limit, if the presence (absence) of an
electron at site $i$ is interpreted as a spin up (down), we immediately see
that this last term is responsible for the presence of geometrical frustration
on the charge degrees of freedom on the triangular lattice, in analogy with magnetic systems.

Besides its fundamental interest {\it per se}, such a two-band description
applies to actual materials with complete $t_{2g}$ and partially filled $e_g$
shells.  In AgNiO$_2$ ($t^6_{2g}e^1{_g}$ configuration, formal
valence Ni$^{3+}$), for example, the orbitals are split between an $e_g$ doublet occupied by
one electron, and a completely filled  $t_{2g}$ triplet that can be neglected
to a first approximation by virtue of the large crystal-field gap of $\sim 2$
eV \cite{VernayPRB04,Kang07,Sorgel07}.
Since the twofold $e_g$ orbitals $| d_{3z^2-r^2} \rangle$ and $| d_{x^2-y^2}
\rangle$ form a pseudospin-$1/2$, we label them, respectively, as $| \uparrow
\rangle $ and $| \downarrow \rangle $ and denote
the corresponding Pauli matrices
by $\tau^i$ with $i=1, 2, 3$.
The transfer integrals along the lattice vectors $\vec{u}_1, \vec{u}_2$, and
$\vec{u}_3= \vec{u}_2- \vec{u}_1$ sketched in Fig. \ref{fig:dos} can be
expressed in terms of two independent parameters $t$ and $t^\prime$
as\cite{SlaterKoster,VernayPRB04,Uchigaito}: 
\begin{equation*}
	t_{\vec{u}_1} = 
   \left (
    \begin{array}{c c}
     t &  0 \\
     0 &  t' \\
    \end{array}
   \right),
~
	t_{\vec{u}_2}  =  
   \left (
    \begin{array}{c c}
     t_2  &  t_3\\
     t_3 & t_4 \\
    \end{array}
   \right),
~
	t_{\vec{u}_3}  =  
   \left (
    \begin{array}{c c}
     t_2 &  \textrm{-}t_3 \\
     \textrm{-}t_3 &t_4
    \end{array}
   \right)
\end{equation*}
with $t_2= (t + 3 t')/4$, $t_3=\sqrt{3}(t - t')/4$ and $t_4=(3 t + t' )/4$.  In
full generality we take $t^\prime/t$ in the interval $[-1,1]$ (all values
outside this interval amount to interchanging orbitals $a$ and $b$) and set
$t$ as the energy unit.  Note that for $t^\prime=t$  the kinetic term reduces
to two independent instances of the triangular isotropic lattice, and the model
becomes analogous to the single-band spinful extended Hubbard model.

The  effective on-site Hubbard repulsion term in Eq. (\ref{eq:H}) describes the
interaction  between electrons on different orbitals.  It obviously has
direct relevance to the study of ferromagnetically ordered states
\cite{Jackeli2008}, in which case only one spin species is present and two
electrons necessarily occupy two different orbitals.  In a more general
context, the present spinless model can be viewed as an approximation to tackle
the strongly interacting  (i.e. strong Hubbard repulsion \textit{and} strong
Hund coupling) limit of the two-band spinful model introduced in
Ref.~[\onlinecite{Ralko15}].  Following standard
notations \cite{Kanamori,Georgesrev}, the
interaction energy of aligned-spin configurations on a site is
$\tilde{U}=U-3J_H$, where $J_H$ is the Hund exchange coupling and $U$ is the
intraorbital Hubbard repulsion.  Such configurations are favored (in other
words,  Hund's coupling favors high-spin states) because all other
configurations have energies $U-2J_H$ and higher.  When $J_H$ approaches $U/3$,
the condition $U-2J_H\gg \tilde{U}$ makes it is possible to restrict the system to states
with aligned spins
since these become energetically more favorable, projecting out all other high-energy configurations.
Spinless fermions constitute a reasonable approximation to this projection for
those aspects of  charge ordering which do not involve the magnetic degrees of
freedom.  Finally, a notable advantage of  the spinless model is that it allows
us to cross-check the mean-field results via the use of 
ED techniques which would be impossible if we considered the full orbital and
spin character in the model.

We mention here that although related multiband models for charge and orbital
ordering on the triangular lattice have been studied in recent years,   how
multiorbital kinetic terms affect frustrated charge ordering (in particular
the pinball liquid) remains an open question.  Vernay \textit{et al.}
\cite{VernayPRB04}, for example, studied the evolution of orbital ordering as a
function of $t^\prime/t$ in a spinful model via both mean-field and exact
diagonalization, but they  did not consider the charge ordering induced by the
intersite repulsion $V$.  Uchigaito et al. \cite{Uchigaito} performed a
mean-field analysis of the effects of both the Hund and Jahn-Teller couplings
on the electron ordering as a function of $t^\prime/t$; however, because the
repulsion $V$ was not included in the model, no pinball-liquid phase was found
at realistic values of the Hubbard repulsion $U$, which is at odds with the experimental
observations in AgNiO$_2$.  The question has also been addressed from an
\textit{ab initio} point of view  \cite{Mazin}, including both local and
nonlocal interaction effects, but without providing systematic studies as a
function of the microscopic Hamiltonian parameters.  Finally, multiband
effects on charge ordering on the triangular lattice have been studied via both
UHF and dynamical mean-field theory in Ref.~[\onlinecite{Ralko15}], but only the
fully isotropic limit $t^\prime=t$ was explored.

\subsection{Methods}
\label{sec:methods}

To solve Eq. (\ref{eq:H}) we have used the UHF and ED methods, which are both
defined on the site basis.  There are two types of clusters with periodic
boundary conditions on the triangular geometry which respect all the symmetries
of the infinite lattice (translations and point-group symmetries). In terms of
the basis vectors $\vec{u}_1$ and $\vec{u}_2$, they are defined by two vectors,
\begin{eqnarray*}
\vec{T}_1 &=& l   \vec{u}_1 + m \vec{u}_2 \\
\vec{T}_2 &=& - m \vec{u}_1 + (l + m) \vec{u}_2 ,
\end{eqnarray*}
with $l$ or $m=0$ for {\it regular} clusters (considered here for the UHF
computations) and $l=m$ for {\it tilted} ones used for the ED. The regular
clusters  have a number of sites $N = l^2$, while it is $N = 3 l^2$ for the
tilted lattices. In order to allow all relevant symmetry breakings, we have
considered lattices for which the number of sites is always a multiple of $12$.

In UHF, the interaction terms $n_{i \uparrow} n_{j \downarrow}$ are decoupled
as a Hartree term  $\langle n_{i \uparrow}\rangle n_{j \downarrow} +  n_{i
\uparrow} \langle n_{j \downarrow} \rangle - \langle n_{i \uparrow} \rangle
\langle n_{j \downarrow} \rangle $ and a Fock term $ \langle d_{i
\uparrow}^\dagger d_{j \downarrow} \rangle d_{j \downarrow}^\dagger d_{i
\uparrow} + d_{i \uparrow}^\dagger d_{j \downarrow} \langle d_{j
\downarrow}^\dagger d_{i \uparrow} \rangle - \langle d_{i \uparrow}^\dagger
d_{j \downarrow} \rangle \langle d_{j \downarrow}^\dagger d_{i \uparrow}
\rangle $.
The sets $ \langle n_{i \tau} \rangle $ and $\langle d_{i \tau}^\dagger
d_{j \tau'} \rangle $, for $i$ and $j\in [1, N]$,
are computed from the wave functions in the single-electron basis.  A
self-consistent loop is performed starting from different initial trial states
(homogeneous, random, charge and/or orbitally ordered) until a fixed point is
reached. The result with the lowest energy is selected among the converged
states.
The method is free from local constraints and \textit{ad hoc} symmetrizations of the
solution; that is, no particular form of the ground state is assumed. This allows
us to obtain the most general ordered states of the model in an unbiased way and
has been proven very successful in predicting novel phases in related models
\cite{Ferhat,Ralko15}.

Also due to the absence of constraints, the resulting mean-field Hamiltonian
does not necessarily commute with $\tau^2$ and $\tau^z$, which are, respectively, the total
orbital pseudospin operator and its $z$ component. 
We notice that the Fock terms of the Hubbard interaction can be recast in terms
of the ladder pseudospin operators $\tau^\pm$ as  $ \langle \tau_i^+ \rangle
\tau_i^-  + \langle \tau_i^- \rangle  \tau_i^+ $, allowing for the extraction of
the average orbital components $\langle \tau^x \rangle $, $\langle \tau^y
\rangle $, and  $\langle \tau^z \rangle $. This is particularly useful in order
to characterize, in addition to the charge symmetry breaking, solutions having
an orbital order such as the $120^{\degree}$ phase already observed in the Hubbard
model at large $U$ \cite{Krishna1,Krishna2}.
To avoid spurious solutions and to obtain smooth convergences, we have employed
here a finite-temperature version of the UHF, checking for all our results that
the low-temperature regime was reached and no thermal fluctuations 
remained.  Typically, this is always achieved for an inverse temperature
$\beta = 1/50 t$.
Finally, we have performed a systematic size-scaling analysis for clusters up to
72$\times$72 sites in order to identify
transition lines in the thermodynamic limit.

For the ED calculations, we have taken the largest available cluster fulfilling
all the symmetry requirements mentioned above, namely, the 12-site tilted
cluster ($l = m = 2$, as depicted in Fig. \ref{fig:dos}). Since the Hamiltonian
does not preserve the orbital flavor, one has to consider, for a fixed number
of particles, all the orbital sectors. In addition, due to the structure of the
hopping matrix, as soon as $t'\ne t$, the three directions of the lattice
become inequivalent hence breaking the $C_6v$ point-group symmetry. The system
can thus be block-diagonalized simultaneously for $k$ points compatible with the
remaining $C_2$ symmetries, which contain only two generators, $Id$ and $R_\pi$,
the rotation of an angle of $\pi$ around the $z$ axis. This is true for only points
$\Gamma$ and  $M$; $K$ and $K/2$ are not compatible with such symmetries
(see the corresponding Brillouin zone in Fig. \ref{fig:dos}).  In a given $k$
sector, the ground state (GS) can then be either symmetric (referred to as $A$) or antisymmetric
($A_p$) against the $C_2$ symmetries. This will be used in the interpretation
of the ED phase diagram.

For both UHF and ED we have identified the different phases via the following
quantities: (i) the charge-charge correlations $C(k) = (1/N)\langle \psi_0 |
\rho( - k ) \rho( k ) | \psi_0 \rangle$, with $\rho(k)$ being the Fourier transform of
the total on-site density operator $n_{i\uparrow} + n_{i\downarrow}$, (ii) the
orbital-orbital correlations $T(k) = (1/N)\langle \psi_0 | \tau^z( - k )
\tau^z( k ) | \psi_0 \rangle$, where $\tau^z$ is the $z$ component of the
pseudospin-orbital operator, and (iii) the average double occupation $D =
(1/N)\langle \psi_0 | \sum_i n_{i \uparrow } n_{i \downarrow } |\psi_0 \rangle
$.  Finally, we  also computed the spectral function   
\begin{eqnarray}
A_\tau(k,\omega) = \frac{-1}{\pi} \textrm{Im} \langle \psi_0 | d_{k\tau}
\frac{1}{\omega - H + i \eta } d^\dagger_{k\tau} | \psi_0 \rangle
\label{eq:akw}
\end{eqnarray}
within the UHF approach to analyze the reconstruction of the band structures
and Fermi surfaces.

\section{Results}

\subsection{Noninteracting limit}
\label{sec:nil}

Figure \ref{fig:dos} reports the evolution of the non-interacting density of
states (DOS) as a function of $t^\prime/t$. For $t^\prime/t=1$, the model
reduces to two independent instances of the triangular lattice. As soon as
$t^\prime\neq t$, however, the electronic dispersion separates into two
nondegenerate bands. Correspondingly,  the logarithmic singularity in the DOS
of  the triangular lattice splits into two peaks, and Dirac cones appear in the
band structure at the $K$ points.  Although the DOS never vanishes at the Dirac
points, due to the simultaneous presence of other bands at the same energy,
these can be identified in Fig. \ref{fig:dos} by the kinks located between the
two Van Hove singularities.  A Lifshitz transition occurs for $t'\simeq 0.43t$
where the lowest of the two peaks
in the DOS crosses the Fermi energy, and one of the two bands changes character
from electronlike to holelike.  The band structure eventually becomes
particle-hole symmetric for $t^\prime=-t$, in which case the Dirac point falls
at the band center. In this case, the system is a semimetal with two parabolic
bands touching at the $\Gamma$ point and Dirac cones at the $K$ points.

\subsection{Unrestricted Hartree-Fock}

\begin{figure}[h]
   \centering 
    \includegraphics[width=8.5cm]{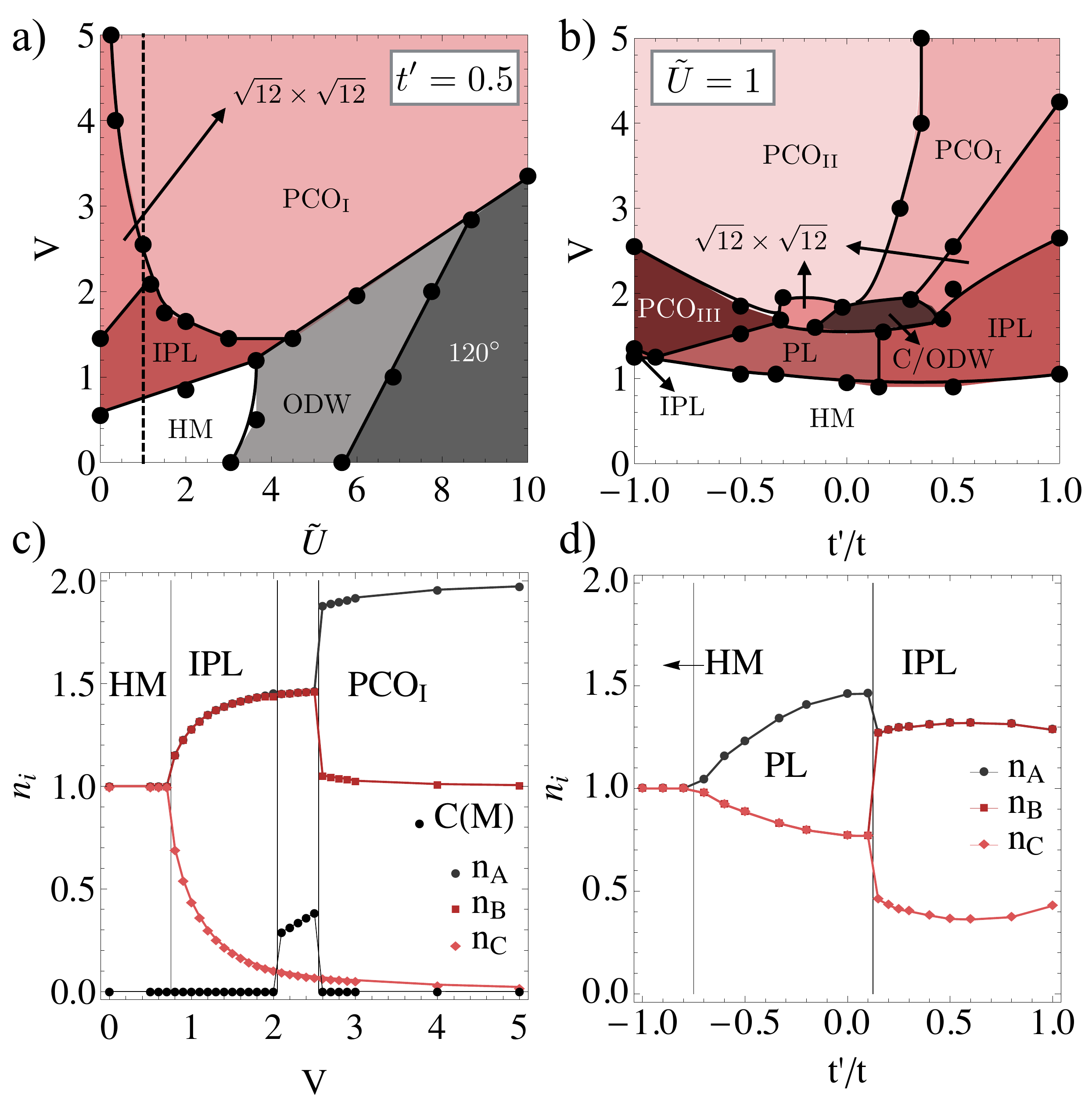}
   \caption{ 
(a) Phase diagram obtained from UHF at fixed  $t' =0.5t$.  The phases denoted
by HM and IPL are metallic, and all the other phases are insulating. IPL
is charge ordered, PCO$_\textrm{I}$ and $\sqrt{12}\times\sqrt{12}$ are charge and
orbitally ordered, and ODW and 120$^{\degree}$ are orbitally ordered. (b) Phase diagram at
fixed  $\tilde{U} = 1$ as a function of $t'/t$. HM, PL, IPL, and
PCO$_\textrm{III}$ are metallic; all phases except HM are charge ordered. (c)
Evolution of the three sublattice densities $n_A$, $n_B$, $n_C$ and charge
correlations $C(M)$ along the constant $\tilde{U}$ line shown in (a)
[the scale of $C(M)$ has been multiplied by $10$].
For the $\sqrt{12} \times \sqrt{12}$ phase, $n_A$, $n_B$, $n_C$ are defined as
the averages over the four nonequivalent sublattices in the new unit cell. (d)
Sublattice densities as a function of $t'/t$ for fixed $\tilde{U}=1$ and
$V=1.1$.
}
 \label{fig:PD}
 \end{figure}

The phase diagram obtained in the $(\tilde{U},V)$ plane for $t^\prime/t=0.5$
from UHF is reported in Fig.  \ref{fig:PD}(a). We choose this representative
value of $t'/t$ because the corresponding noninteracting DOS qualitatively
reproduces the main features calculated by DFT-LDA for AgNiO$_2$ \cite{Sorgel07}.
To explore all the possible regimes of the model, we also display in Fig.
\ref{fig:PD}(b) the phase diagram  in the $(t',V)$ plane at a fixed value of
$\tilde{U}=1$.  We recall here that such a moderate $\tilde{U}$ actually stands
for a local Hubbard repulsion $U$ which is considerably larger than this value.
As discussed after Eq. (\ref{eq:H}), $\tilde{U}$ is an effective parameter
which is related to the microscopic interaction parameters in the $d$-electron
manifold via the equality $\tilde{U}=U-3J_H$.  For a choice of the Hund
coupling $J_H=0.25U$, for example, the value $\tilde{U}=1$ corresponds to
$U=4$, a value appropriate to moderately correlated materials.
  
As we show below, a rich variety of phases is found and can be classified
into charge ordered (at large $V$) and charge homogeneous (at low $V$). 
\begin{figure}[h]
   \centering 
    \includegraphics[width=0.45\textwidth]{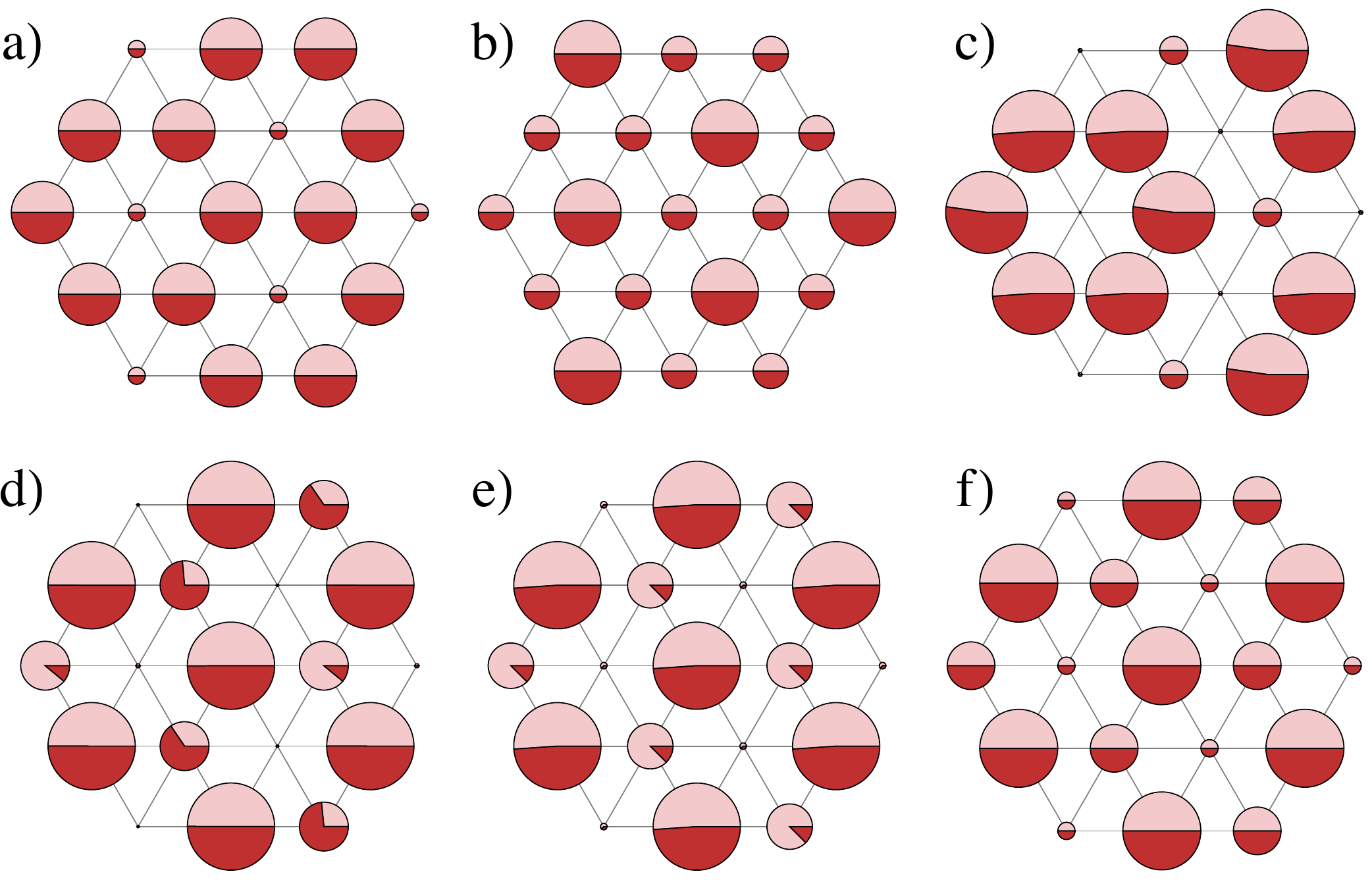}
   \caption{ Typical chargei- and orbital-density snapshots in the different
charge-ordered phases realized in the model: (a) IPL, (b) PL,  (c)
$\sqrt{12}\times \sqrt{12}$, (d) PCO$_{\rm I}$, (e) PCO$_{\rm II}$, and  (f)
PCO$_{\rm III}$.  The radius of the disks is proportional to the charge density
on the sites, and the light and dark fillings correspond to the partial orbital
densities. } 
 \label{fig:snap}
 \end{figure}
Among the latter, several types have been identified in the model, whose charge
and orbital density patterns are depicted  in Fig. \ref{fig:snap}.  These are
denoted as inverse pinball liquid (IPL), PL, pinball
charge order (PCO), and the $\sqrt{12} \times \sqrt{12}$ droplet phase.  Supplementing
the real-space snapshots, Fig. \ref{fig:PD}(c) conveniently reports the
evolution of the local charge densities as a function of $V$ along the vertical
line shown in Fig. \ref{fig:PD}(a) corresponding to $\tilde{U}=1$. An analogous
scan is presented in Fig.  \ref{fig:PD}(d) for the $t'$ dependence at fixed
$V=1.1$.

\begin{figure*}[t!]
   \centering 
    \includegraphics[width=12cm]{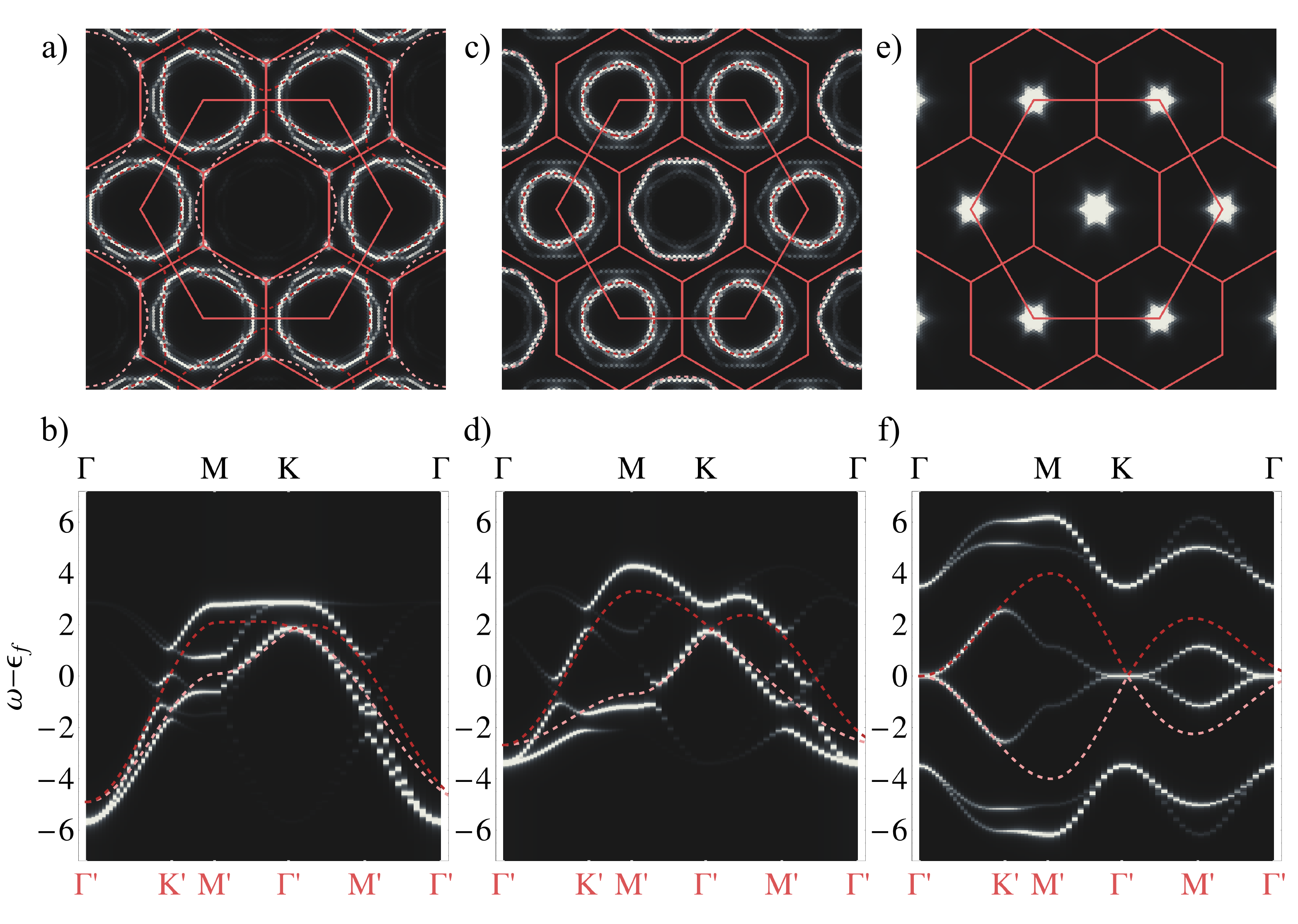}
   \caption{
Fermi surfaces and spectral functions for the threefold metallic
charge-ordered phases, obtained from Eq.  (\ref{eq:akw}) with a Lorentzian broadening
$\eta=0.02$ on 72$\times$72 site lattices: (a) and (b) IPL
($V=0.9,\tilde{U}=1,t'=0.5$), (c) and (d) PL ($V=1.1,\tilde{U}=1,t'=0$), and (e)
and (f) PCO$_{\textrm{III}}$ ($V=2,\tilde{U}=1,t'=-1$).  The solid lines in the top
panels show the original and reduced Brillouin zone, whose symmetry points are
labeled in Fig. \ref{fig:dos}.  The dashed lines are the Fermi surfaces and
band structures in the noninteracting limit.
}
 \label{fig:bands}
 \end{figure*}

\paragraph{Inverse pinball liquid.} \label{sec:IPL} The IPL order develops in
the  whole region $t'/t>0.1$ upon increasing the intersite repulsion $V$ from
the homogeneous metal phase (HM) for sufficiently low values of $\tilde{U}$
[Figs. \ref{fig:PD}(b) and \ref{fig:PD}(c)]. This phase exhibits charge order with a
three-sublattice structure (A, B, C).  The charge density separates into charge-rich
sites forming a honeycomb lattice ({\it balls}, density $n_A=n_B=1+\delta$,
with $\delta$ being the charge disproportionation) and charge-poor sites ({\it pins},
density $n_C=1-2\delta$) located on the remaining triangular sublattice (we
conventionally label the sublattice densities in descending order,
$n_A>n_B>n_C$).  The charge disproportionation  progressively increases with
$V$ towards the maximum allowed value $\delta=1/2$ corresponding to a fully
depleted charge-poor sublattice ($n_C=0$ and, correspondingly, $n_A=n_B=3/2$).
There is no orbital polarization in either the charge-rich or charge-poor sublattices.

The IPL is metallic due to the presence of itinerant carriers (balls) on the
charge-rich honeycomb network. Figure \ref{fig:bands}(a)  illustrates the Fermi
surface (FS) obtained in this phase ($V=0.9$, $\tilde{U}=1$, $t'=0.5$), which
clearly shows the existence of holelike carriers around the $K$ points of the
original Brillouin zone (large hexagon), resulting from the folding of one of
the noninteracting bands (shown as dashed lines).  Small  pockets can also be
seen around the corners of the reduced Brillouin zone (small hexagons).  These
are remnants of the second band of the noninteracting system, which at this
value of $t'$ crosses the reduced Brillouin zone very close to its corners
(denoted as $K'$ points) and therefore folds into closed pockets of trigonal
shape.  The origin of the large hole FS  and the small trigonal pockets can
also be clearly seen by comparing the spectral function $A(k,\omega)$
illustrated in Fig. \ref{fig:bands}(b) with the dispersion of the two
noninteracting bands (dashed lines).

\paragraph{Pinball liquid.}
\label{sec:PL}
When $t'/t<0.1$, the IPL $\frac{3}{2}$-$\frac{3}{2}$-0 charge pattern shown in
Fig. \ref{fig:snap}(a) is replaced by the PL, with a
2-$\frac{1}{2}$-$\frac{1}{2}$ pattern as illustrated in Fig. \ref{fig:snap}(b).
This phase is dual to the IPL in the sense that  the roles of the charge-rich
and charge-poor sublattices are interchanged.  We identify this phase  with the
original PL of Hotta and coworkers \cite{Hotta06} because, from the point of
view of the electronic densities ($n_A=2$, $n_B=n_C=1/2$), it can be viewed as
two realizations of the PL phase found at quarter filling, one per each
orbital character (the PL at quarter filling has $n_A=1$, $n_B=n_C=1/4$).
Because of orbital-orbital interactions, however, the two realizations are
not independent, and the present PL can occur for only small or moderate values
of $\tilde{U}\lesssim 5$. This can be contrasted with the quarter-filled case,
where a strong Hubbard term is required to stabilize the PL \cite{Cano10,Cano11,Merino13}.

The phase diagram in Fig. \ref{fig:PD}(b) shows that the selection between the
PL order and its dual IPL  is entirely governed by kinetic effects.  This can
be understood by observing that the electrostatic energies of the PL and IPL are
formally equal and do not depend on the sign of the charge disproportionation
$\delta$, $E_{IPL}=E_{PL}=3V(1-\delta^2)$.

Similar to the IPL, metallic behavior arises in the PL from the motion of
electrons living on the honeycomb network.  At this value of $t'$, the
noninteracting FS (dashed line) is composed of a large electronlike surface
around the $\Gamma$ point and smaller holelike pockets around the $K$ points.
Contrary to the IPL shown previously, however, the overall shape of the FS is
only weakly affected by charge ordering of the PL type because the original Fermi
pockets do not cross the boundaries of the reduced Brillouin zone [Fig.
\ref{fig:bands}(c) and \ref{fig:bands}(d)].

\begin{figure*}[t]
   \centering 
    \includegraphics[width=12cm]{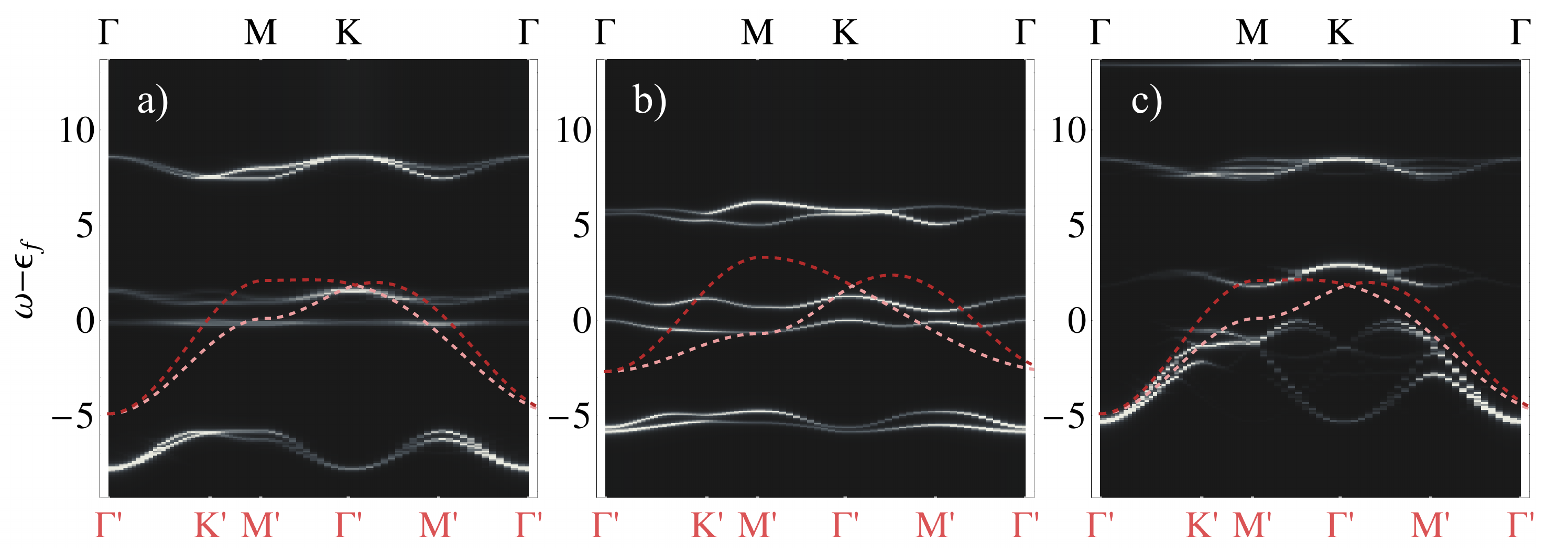}
   \caption{
Spectral functions for the insulating charge-ordered phases: (a)
 PCO$_{\textrm{I}}$ ($t'/t=0.5$, $\tilde{U}=1$, $V=2.6$), (b) 
 PCO$_{\textrm{II}}$  ($t'/t=0$, $\tilde{U}=1$, $V=2$), and (c)
 $\sqrt{12}\times\sqrt{12}$ droplet phase  ($t'/t=0.5$, $\tilde{U}=1$, $V=2.5$).
  The dashed lines represent the band structures in the noninteracting limit.
 }
 \label{fig:bandsins}
 \end{figure*}

\paragraph{Pinball charge order.}
At  large $V$, the system undergoes further charge ordering, stabilizing  a
2-1-0  charge pattern termed pinball charge order [Figs.
\ref{fig:snap}(d)-\ref{fig:snap}(f)].  Simple electrostatic arguments predict that the PCO is
stabilized for $V>\tilde{U}/3$, which nicely agrees with the numerical results
at large $\tilde{U}$ and $V$ [see Fig. \ref{fig:PD}(a)].  The threefold
disproportionation splits the electronic dispersion into three separate bands per
orbital state.  Counting the occupied states leaves us  with a central
half-filled band, which should lead, in principle, to a metallic behavior. We
find instead that  the PCO phase found in the whole region $t'/t\gtrsim 0.1$,
denoted PCO$_\textrm{I}$ in Figs. \ref{fig:PD}(a) and \ref{fig:PD}(b), is insulating
[see the spectral function in Fig. \ref{fig:bandsins}(a)]. This is  ascribed  to the
presence of a spiral $120^{\degree}$ orbital order on the singly occupied $B$ sites,
caused by the local interaction $\tilde{U}$, as seen in the snapshot in Fig.
\ref{fig:snap}(d).  Other orbital orderings on the $B$ sublattice are possible
within the present 2-1-0 charge arrangement upon varying $t'/t$ [Fig.
\ref{fig:PD}(b)].  These are PCO$_\textrm{II}$ for $t'/t\lesssim 0.1$ and
large $V$ [uniformly polarized B sites, Fig.  \ref{fig:snap}(e)] and
PCO$_\textrm{III}$ for $t'/t\lesssim 0.1$ and low $V$ [unpolarized  B sites,
Fig.  \ref{fig:snap}(f)]. PCO$_\textrm{II}$ is also an insulator because
the ferro-orbital polarization is sufficient to split the narrow half-filled band at
the Fermi energy into two separate components [Fig. \ref{fig:bandsins}(b)].
Note that this is the only phase in the whole phase diagram which supports a
nonzero $\langle \tau^z \rangle$.
PCO$_\textrm{III}$ instead has a metallic character.  In the limit $t'=-t$,
shown in  Fig. \ref{fig:bands}(c), it is a semimetal with bands touching at
the $\Gamma$ and $K$ points.

If one interprets as usual the 120$^{\degree}$ ordering as the mean-field equivalent of
a Mott insulating phase, it can be argued that the PCO is a neat example where
charge ordering enhances the effect of Hubbard-type interactions, which happens
because  the reconstruction of the band structure leads to half-filled bands
that are much narrower than in the homogeneous phase, as these rapidly shrink
upon increasing $V$ [Fig. \ref{fig:bandsins}(a) and \ref{fig:bandsins}(b)].  One notable example
of this positive interplay between charge-ordering and the Mott mechanism is
found in the dichalcogenide TaS$_2$ \cite{Perfetti06,Sipos} (although the
charge ordering pattern there has a larger periodicity of 13 sites per cell,
as discussed below). 
\begin{figure*}[!t]
   \centering 
    \includegraphics[width=0.8\textwidth]{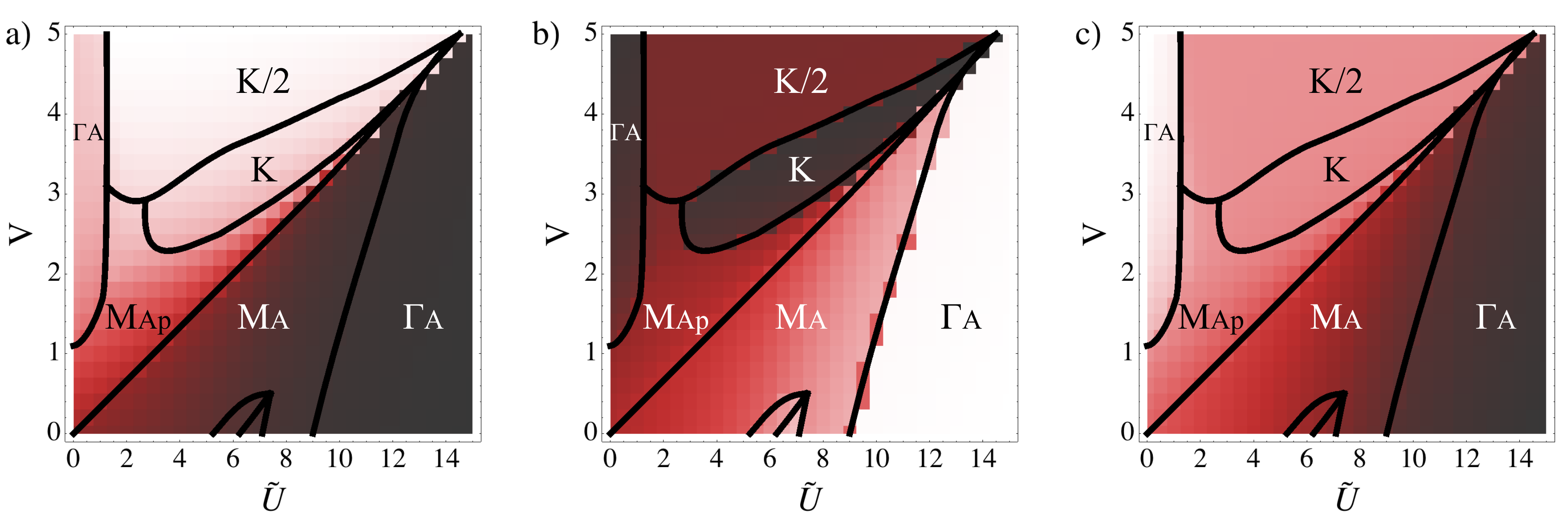}
   \caption{ 
Phase diagrams obtained from ED on a 12-site cluster at fixed  $t' =0.5t$. 
(a) Charge-charge correlation $C(K)$.
(b) Orbital-orbital correlation $T(K)$. 
(c) Expectation value of the double occupation $D$.
The color maps are in arbitrary scale, increasing from dark to bright. On each
panel are shown the ground-state symmetry sectors (translation and point
group) and first-order transition lines (thick lines) reflecting symmetry
breaking driven either by the charges, the orbitals or both. The symmetry sectors in the small dome at low $V$ is respectively $K/2$ and $M_{A_p}$ as $\tilde{U}$ increases.
}
 \label{fig:ED}
 \end{figure*}

\paragraph{$\sqrt{12} \times \sqrt{12}$ droplet phase.}
For $t'/t>0.1$, an additional  phase emerges in the small-$\tilde{U}$ regime,
located between the two charge-ordered phases found at small and large $V$.
This phase, which is stabilized with respect to the PCO phase by purely kinetic
effects, \footnote{It can be shown straightforwardly that the potential energies
of the droplet and PCO phases coincide when $\tilde{U}=0$  (the cost due to $V$
being exactly equal), while the droplet phase is rapidly destabilized by
$\tilde{U}$ due to  the larger double occupation.} has a large periodicity with
12 sites in the unit cell. It is  detected by the coexistence of peaks in the
charge-charge correlation function at the  $K$ points (characterizing threefold
order) plus all the $M$ (whose combination describes a phase with fourfold
symmetry \cite{Tremblay}) and $K/2$ points. The latter are absent in the
phases with pure threefold symmetry.   The $M$-point correlation
 is shown in Fig. \ref{fig:PD}(c).  The
real-space snapshot  in Fig. \ref{fig:snap}(c)  shows that the charges form
disconnected hexagonal droplets, composed of essentially doubly occupied sites.
These are surrounded by sites which are either empty or almost empty.
Correspondingly, this phase has a strongly insulating character due to the
opening of a gap $\propto V$ at the Fermi energy [Fig. \ref{fig:bandsins}(c)].
This phase also displays weak orbital order on the hexagons, signaled by 
nonzero orbital correlations at points $M$ and $K/2$.

We note that the emergence of a charge pattern with a high-order periodicity
such as the one found here is not at all trivial given that the
electron-electron repulsion is 
restricted to only nearest-neighbor sites.
The region in parameter space where the $\sqrt{12} \times \sqrt{12}$ droplet phase
is stabilized actually includes the limit $t'=t$ where  Eq. (\ref{eq:H})
becomes equivalent to the spinful extended Hubbard model, but it was
overlooked in previous studies
\cite{Santoro,Tremblay,Gao,Tocchio14,Hassan,Kanada}.  Interestingly, the
present 12-site period is  very close to the 13-site star of David modulation
found in the triangular dichalcogenide compound  $1T$-TaS$_2$\cite{Sipos}.  The
hexagonal droplets in Fig. \ref{fig:snap}(c) are also similar to those recently
predicted theoretically in the kagome lattice at $n=1/3$ filling
\cite{Ferhat,Pollmann}, but in that case they are a natural consequence of the
larger unit cell of the  underlying lattice (see also  Refs.~[\onlinecite{Koshibae, Peil}]).

Finally, we mention that the $\sqrt{12} \times \sqrt{12}$ phase obtained in the
narrow interval $-1/3 < t'/t \lesssim -0.1$ [see Fig. \ref{fig:PD}(b)]
has the doubly occupied hexagons replaced by empty sites and vice versa  and
can therefore be considered the dual to the droplet phase described above.
The two are separated by a phase with coexisting incommensurate charge and
orbital order, denoted C/ODW.

\paragraph{Homogeneous orbitally ordered phases.}

Several phase transitions are also found within the charge homogeneous region
at low $V$. The system evolves upon increasing the effective local interaction
from a paraorbital metal at small $\tilde{U}$ to a spiral 120$^{\degree}$ orbitally
ordered insulating phase at large $\tilde{U}$ (with the orbital moments
arranged in planes perpendicular to the lattice as soon as $t'\neq t$), going
through an intermediate region with more complex orderings, denoted as ODW in
Fig. \ref{fig:PD}(a).  Although a precise study of this intermediate region is
beyond the scope of the present work, we would like to stress the following
points: (i) Within our real-space unrestricted approach, we have recovered the
two intermediate mean-field phases of the spinful Hubbard model
\cite{Krishna1,Krishna2}, namely an incommensurate orbital (spin) ordered phase
whose wave vector evolves with $\tilde{U}$ and a zigzag stripe compatible with
the points $k=(\pm\pi,0)$. (ii) While previous studies have looked for solutions
breaking the symmetry at a single wave vector, we find solutions compatible
with two or more coexisting $k$ vectors, possibly a mean-field indication of a
tendency to a structureless orbital liquid state.  (iii) Away from the case of
the isotropic Hubbard model,
i.e., as soon as $t' \ne t$
, the zigzag stripe phase seems to disappear, and only the incommensurate
regions with coexisting order survive.
(iv) At $t'=0.5$, the 
critical parameter $\tilde{U}_c$ is decreased by 25\% with respect to the isotropic case $t'=t$, in agreement with the corresponding reduction of the bandwidth.

\subsection{Exact diagonalization}

To ascertain if the variety of phases found at mean-field level in Fig.
\ref{fig:PD}(a) is robust against quantum fluctuations, we have performed a
systematic study of the model equation (\ref{eq:H}) via Lanczos diagonalization on
a 12-site cluster.  Such a cluster is compatible with all the symmetries of the
expected charge-ordered phases except for PCO$_\textrm{I}$, which has a
nine-site unit cell due to the presence of additional 120$^{\degree}$ orbital order. It is
also compatible with the orbital orders found by UHF in the low-$V$ region,
except for the $(\pm\pi,0)$ zigzag stripes and the incommensurate stripes
characteristic of intermediate $\tilde{U}$.  

Figure  \ref{fig:ED} shows the phase diagram obtained in the $(U,V)$ plane for
$t' = 0.5 t$ by combining the symmetry character of the ground state for the
charge-charge [Fig. \ref{fig:ED}(a)] and the orbital-orbital [Fig. \ref{fig:ED}(b)] correlations at the corner of the
Brillouin zone ($K$) and  the average double occupation [Fig. \ref{fig:ED}(c)].  Our ED results
for small clusters confirm that the very rich physical picture created by the
interplay between charge and orbital degrees of freedom persists  even beyond
the mean-field level.  A large number of domains with different symmetries are
obtained, separated by first-order transitions,  as displayed in each panel of
Fig.  \ref{fig:ED};
these domains can be associated with all of the different phases
found in Fig.  \ref{fig:PD}(a).

\textit{a. Charge-ordered phases.}
The buildup of $K$-point charge correlations in Fig.  \ref{fig:ED}(a), which is
expected in all the charge-ordered phases described in Sec. IIIB, shows
remarkable agreement with the UHF charge-ordering transition lines reported in
Fig. \ref{fig:PD}(a) and asymptotically follows the analytical prediction
$V_c=\tilde{U}/3$.  We  provide in Table \ref{tab:ts} some representative
values of $C(k)$ at points $K$ and $M$ for the different charge-ordered phases,
as computed numerically by both UHF and ED and analytically on ideal electrostatic patterns. As one can see, a quantitative
agreement is found between the two numerical methods, which allows for direct
identification of the  mean-field phases in the ED results.  From Fig.
\ref{fig:ED} and Table  \ref{tab:ts}, we associate the different charge-ordered
phases as follows: $K/2$ and $K$ $\to$ PCO, $M_{A_p}\to$ PL, and $\Gamma_A\to$
$\sqrt{12}\times\sqrt{12}$.

\begin{table}[!t]
\begin{ruledtabular}
\begin{tabular}{@{}l | l | l | l | l@{}} 
			&				&	PCO		&	IPL/PL			& $\sqrt{12}\times \sqrt{12}$ \\
\hline
Analytical	&	$K$    	&     	1/3 	&	1/4 	    &	1/4 		\\
			&	$M$		&		0		&		0		&	1/9 	\\
\hline                
UHF			&	$K$		&   0.302  		&	0.121		&	0.227	\\		
			&	$M$		&	0	        &	0	        &	0.065	\\
\hline                
ED			&	$K$		&	0.308		&	0.157		&	0.262	\\
			&	$M$		&	0.006		&	0.055		&	0.063	\\
\end{tabular}
\end{ruledtabular}
\caption
{
Comparison of the charge-charge correlation $C(k)/N$ for the high symmetry
points $K$ and $M$ in the different charge-ordered phases,  computed
analytically on the ideal patterns, via UHF and ED. The values for PCO, IPL/PL,
and  $\sqrt{12}\times \sqrt{12}$  correspond to $t'/t=0.5$ and $(\tilde{U},V)$
of (6,4), (1,1.2), and (0,3), respectively.
}
\label{tab:ts}
\end{table}

First, we identify both the $K/2$ and $K$ ground states with the broad PCO
region of Fig. \ref{fig:PD}(a).
We attribute the additional transition seen here, which is absent in UHF,  to
the fact that the ninefold orbital order present in the PCO$_\textrm{I}$ phase is
not compatible with the 12-site cluster, so that different orderings are
stabilized instead. This may due to either the exact treatment of the correlations or the size of the lattice.
Accordingly,  this change of
symmetry is not observable in $C(K)$, but it is clearly seen in $T(K)$  in Fig.
\ref{fig:ED}(b).

Second, by looking at the charge correlations in real space (not shown), it is
possible to associate the $M_{A_p}$ ground state with PL order.  The PL is
found here instead of the  IPL expected from UHF because the considered
cluster is too small to capture the subtle kinetic effects which distinguish
between these two phases.  This is confirmed by the fact that the PL is also
selected in the UHF solution when a 12-site cluster is considered, as we have
checked.

Third, the $\sqrt{12}\times \sqrt{12}$ droplet phase found in the mean field can be
associated with the  $\Gamma_A$ ground state at low $\tilde{U}$ in Fig.
\ref{fig:ED}. This  phase has a charge signature corresponding to a mixture
of the high-symmetry $k$ vectors $K$,  $M$, and $K/2$, with dominant weight on
the first two, which agrees with the UHF and analytical results (see Table
\ref{tab:ts}).  Also, it is in this phase that we find the strongest double
occupancy [Fig.  \ref{fig:ED}(c)], which corresponds to the doubly occupied sites on
the hexagons in Fig. \ref{fig:snap}(c).  Our ED  results confirm the finding
that the $\sqrt{12} \times \sqrt{12}$ droplet phase remains stable in the
spinful extended Hubbard model ($t'=t$).

\textit{b. Orbitally ordered phases.}
The different phases found in UHF upon increasing $\tilde{U}$ at low $V$ also
have their direct analogs in ED.  We associate  the spiral $120^{\degree}$ phase
obtained in UHF with the $\Gamma_A$ phase at large $\tilde{U}$ and small $V$:
here the double occupation is strongly suppressed, indicating a Mott insulating
state with large threefold orbital correlations compatible with such an order
\cite{Capone,Senechal}. We note that the critical value for the Mott
transition, $U_c\approx 9t$, is smaller than the value $U_c\approx 12t$
reported from analogous ED calculations in the isotropic case \cite{Capone},
which can be understood by observing that the bandwidth is reduced by roughly
$25\%$ for the considered $t'=0.5t$ (Fig.\ref{fig:dos}).  We also observe a
small dome in the middle of the $M_A$ phase, which we tentatively associate
with the ODW orders obtained by UHF on larger systems as suggested by the
signatures in $T(K)$. In this dome, two distinct symmetry sectors corresponding
to two different phases are encountered as $\tilde{U}$ increases, $K/2$ and
$M_{A_p}$, respectively.

\section{Conclusions}

We have considered a minimal electronic model which describes the interplay
between frustrated electron-electron interactions and multiorbital effects on
the half-filled triangular lattice.  Our results based on the combination of a
fully unrestricted Hartree-Fock method, which provides an accurate description
of the multiband kinetic properties, and exact diagonalization on small
clusters, which properly takes into account interorbital and intersite
correlation effects, reveal a very rich phase diagram. A number of original
charge-ordered and orbitally ordered phases are displayed, whose occurrence can be
tuned  by varying the multiorbital band structure parameters and which could
be relevant to a variety of $d$-electron systems on the triangular lattice.
These include  threefold metallic charge-ordered phases such as the pinball
liquid, which was originally predicted to occur in quarter-filled lattices and
is shown here to be stable at half filling, as well as its dual, the inverse
pinball liquid.  Both these phases
could be relevant to AgNiO$_2$, where a robust threefold charge-ordered
metal has been experimentally observed \cite{Coldea07,Coldea11}, which we
therefore associate with a pinball state (see also
Ref.~[\onlinecite{Ralko15}]).  An original insulating droplet phase with a
large $\sqrt{12} \times \sqrt{12}$ periodicity is also obtained here, which was
overlooked in previous studies of the extended Hubbard model and which could
be closely related to the star of David charge-ordered phase of the triangular
dichalcogenide compound $1T$-TaS$_2$\cite{Sipos}.  The insulating threefold PCO
phase obtained here could also find a possible experimental realization, as
suggested in triangular absorbate layers \cite{Cortes13}. All these possible
connections with the experiments emphasize the general nature of our study and
should motivate further investigations in stabilizing original charge orders
driven by frustrated electronic interactions and kinetic effects.

Finally, we note that in this paper we have restricted our study to a perfectly
stoichiometric case where the $e_g$ doublet is initially quarter filled,
corresponding  to an average valence Ni$^{3+}$ in AgNiO$_2$.  It has  been
proposed, however, that the filling of the $e_g$ levels in nickelates might actually
differ from this value \cite{Demourgues,Mizokawa}, as extra electrons could be
transferred from the oxygen 2$p$ orbitals provided that the electrostatic cost
$\tilde{U}=U-3J_H$ on the Ni sites is sufficiently low \cite{Mazin,Subedi}.
The very observation of a pinball state in AgNiO$_2$ indicates that the
effective interaction $\tilde{U}$ is low in this material [see Fig. 2(a)], and
the existence of a ``negative charge transfer'' from the oxygen atoms is indeed
compatible with the estimates of Ref.~[\onlinecite{Coldea11}], which indicate a reduced
average valence of 2.85+ per Ni.  Studying the present model away from the
perfectly quarter-filled configuration will certainly be of interest in view of
these
considerations.

\section*{Acknowledgments.} The authors thank J. Merino and L. de' Medici for
useful discussions. This work is supported by the French National Research
Agency through Grants No.  ANR-12-JS04-0003-01 SUBRISSYME and No.
ANR-2010-BLANC-0406-0 NQPTP.

\end{document}